\magnification=\magstep1
\parindent=0pt
\parskip=6pt
\openup 1\jot
\font\titlefont = cmbx10 scaled \magstep3
\font\authorfont = cmbx10 scaled \magstep1
\input amssym.def
\input amssym.tex

\rightline {DAMTP 96-111}
\vskip 20pt
\centerline {\titlefont First Order Vortex Dynamics}
\vskip 20pt
\centerline{{\authorfont N.S. Manton}\footnote {${}^*$} {email:
N.S.Manton@damtp.cam.ac.uk}}
\vskip 15pt
\centerline{\it Department of Applied Mathematics and Theoretical
Physics} 
\centerline{\it University of Cambridge} 
\centerline{\it Silver Street, Cambridge CB3 9EW, England}
\vskip 30pt

{\bf Abstract}
\vskip 5pt
A non-dissipative model for vortex motion in thin
superconductors is considered.  
The Lagrangian is a Galilean invariant version of the Ginzburg--Landau
model for time-dependent fields, with kinetic terms linear in the
first time derivatives of the fields.  It
is shown how, for certain values of the coupling constants, 
the field dynamics can be
reduced to first order differential equations for the vortex positions.  Two
vortices circle around one another at constant speed and separation in
this model.

\vfill\eject
{\bf 1. Introduction}

Magnetic flux penetrates a Type II superconductor in the form of
vortices [1], and recently it has become possible to produce images of
vortices sufficiently rapidly that their motion can be observed
directly [2].
In the Ginzburg--Landau theory of
superconductivity, a charged scalar field representing the electron-pair
condensate is coupled to the electromagnetic field.  The basic vortex
solution, discovered by Abrikosov [3], is a localised magnetic flux
tube surrounded by a circulating supercurrent.

The Ginzburg--Landau potential energy functional contains only one 
dimensionless coupling constant  $\lambda$.  The 
value $\lambda = 1$ (in our units) is mathematically particularly interesting,
because in this case there are no forces between static vortices, 
and there is a continuous family of static multivortex solutions. A
Type II superconductor is modelled by $\lambda > 1$.   In this case,
the potential energy of a two-vortex configuration
decreases as the separation increases, in other words, vortices repel
[4].  However, there are several possibilities for how the vortices
might move, depending on the nature of the dynamical equations for the
fields.  Let us ignore pinning, which tends to prevent vortex motion at
all. The
first possibility is that the vortex acceleration is proportional to the force
acting.  This is what occurs in the relativistic generalisation of the
Ginzburg--Landau model, known as the Abelian Higgs model. Relativistic
vortices may be interpreted as a solitonic version of
fundamental strings [5], or as strings joining confined quarks, or as
cosmic strings produced at a phase transition early in the universe's
history [6]. The second possibility is that the vortex velocity is
proportional to the force.  This is modelled by dissipative equations
involving the first time derivatives of the fields [7]. Recently, one
version of such equations, the Ginzburg--Landau gradient flow
equations, have been analysed in detail [8].  The third
possibility is that the vortex motion is at right angles to the force,
so two vortices circulate around each other, as in a fluid, and there
is no dissipation [9].
In real superconductors, vortex motion is usually dissipative, 
but at very low temperatures, it has been argued that the
third type of motion would occur [10].

The response of vortices to an applied electric ``transport'' current,
perpendicular
to the vortex cores, can distinguish the second and third types of
motion.  If the vortices move at right angles to the current, then the
dynamics is dissipative, but if they are carried along by the current
(again, as in a fluid) then the dynamics is non-dissipative.  To see
this, note that a moving vortex has an electric field in its core,
perpendicular to both the velocity and the direction of the magnetic
flux.  Also, part of the applied current penetrates the vortex core,
where it becomes a normal electric current. There is
dissipation when the current and
electric field in the core are parallel, but not when they are perpendicular.

The purpose of this paper is to analyse a model for the field
dynamics in a thin, essentially two-dimensional superconductor, and to
show that it leads to vortex motion 
of the third type.  The field equations are obtained from a Lagrangian whose
kinetic terms are linear in the first time derivatives of the fields
and whose potential part is the usual Ginzburg--Landau energy.  This
Lagrangian and its associated field equations are given in section 2.
The Lagrangian is Galilean invariant, so we can see precisely how
vortices respond to a transport current.

Section 3 is a review of the static vortex solutions of the
Ginzburg--Landau theory, focussing especially on the manifold $M^n$, the
$2n$-dimensional parameter space of static $n$-vortex solutions which exist at
the critical coupling $\lambda = 1$. These solutions are also present
in the model considered here, at special values of the couplings.

Section 4 treats the case where $\lambda$ is close to $1$, and where
$n$-vortex motion can be well approximated 
by a motion on $M^n$.   This adiabatic approximation
assumes that at each instant the field is a static solution, but that the
parameters of the static solution
(i.e. the vortex positions) slowly vary with time.  It is shown how
the kinetic energy and potential energy of the field Lagrangian can be reduced
to give a finite-dimensional dynamical system on $M^n$.  The kinetic
energy can be expressed in terms of local data associated with each of
the vortices, and although an
explicit form for this is not obtained, some
conclusions can be drawn.  In section 5 
the equations of vortex motion are derived from the reduced Lagrangian, and
it is shown that two vortices circle around one other.

The method used here for reducing the field dynamics to a particle dynamics for
vortices closely follows the analysis of slow vortex motion in the
Abelian Higgs model with $\lambda = 1$, as carried out by
Samols [11].  There the reduced system's kinetic energy expression
involves a Riemannian metric on $M^n$, the potential energy is a
constant, and the vortex dynamics is
modelled by geodesic motion on $M^n$ [12]. The geodesic motion is
modified by the effect of a potential energy varying over $M^n$ if 
$\lambda$ is not
exactly unity.  This was studied by Shah [13]. Our reduced system has
the same potential energy, and a kinetic energy which involves a
connection or gauge potential on $M^n$ that depends on the same data
as Samols' metric.

{\bf 2. The Schr\"odinger--Chern--Simons Lagrangian}

Let us consider a two-dimensional superconductor, with translational
symmetry.  A thin film with no defects might be close to this
idealization. There is a complex scalar field $\phi$, representing the
electron-pair condensate, and a gauge potential with time component
$a_t$ and spatial components $a_i : i = 1,2$.
The magnetic field is $B = \partial_1 a_2 - \partial_2
a_1$ and the electric field $E_i = \partial_i a_t - \dot a_i$. 
(An overdot denotes ${\partial \over \partial t}$.)  
Units are fixed so
that $\phi$ has magnitude $1$ in the condensed phase (vacuum), and its
covariant time and space derivatives are 
$D_t\phi = \partial_t\phi - i a_t \phi$ and
$D_i\phi = \partial_i\phi - i a_i \phi$.

The Lagrangian that we shall consider, $L$, is an
expression involving no higher than
the first power of time derivatives. Explicitly,
$$
\eqalign{
L =  \int \Bigl( \gamma {i \over 2} 
&\left ( \phi^* D_t \phi - \phi (D_t \phi)^* \right ) 
+ \mu \left ( B {a_t} + E_2 a_1 - E_1 a_2 \right )
- \gamma a_t \cr
&- {1\over 2} B^2 - {1\over 2} (D_i \phi)^* D_i \phi - {\lambda\over
8} (1 -  |\phi |^2)^2 - a_i J^T_i \Bigr) d^2x  \ , \cr }
\eqno(2.1)
$$
where ${1\over 2} B^2 + {1\over 2} (D_i \phi)^* D_i \phi + {\lambda\over
8} (1 -  |\phi |^2)^2$ is the standard
Ginzburg--Landau field energy density for static fields.  
Since there is no relativistic invariance, the summation
convention is used only in the two space dimensions.  The term with coefficient
$\mu$ is the Chern--Simons density for the gauge field.  $\gamma , \mu$
and $\lambda$ are real constants, with $\lambda$ positive.  $J_i^T$ is
the transport current, that is, a constant vector in the plane of the
superconductor.  This Lagrangian is hardly original.  The
scalar field part of the Lagrangian coupled to the Chern--Simons term
has appeared in the theory of Chern--Simons vortices, developed by
Jackiw, Pi and others [14].
The inclusion of the $\gamma a_t$ term was advocated by Barashenkov and
Harin to allow the possibility of a condensate $(|\phi | = 1)$ at
infinity [15].  Somewhat original is that the ${1\over 2} B^2$
term is included, but not the ${1\over 2} E_i E_i$ term usually present in the
Maxwell Lagrangian.  This is justified because there is no
relativistic invariance in the rest of the Lagrangian, and it is desired to
have only first time derivatives in the field equations.

The field equations are
$$
\eqalignno{
i \gamma D_t \phi &= - {1\over 2} D_i D_i \phi - {\lambda \over 4} (1 -
|\phi |^2) \phi &(2.2) \cr
\epsilon_{ij} \partial_j B &= J_i^S - J_i^T + 2\mu \epsilon_{ij} E_j &(2.3)\cr
2\mu B &= \gamma (1- |\phi |^2) \ , &(2.4) \cr
}
$$
obtained by varying with respect to $\phi^*$, $a_i$ and $a_t$
respectively. $J_i^S$ is the supercurrent, defined by
$$
J_i^S = -{i\over 2} \left ( \phi^* D_i \phi - \phi (D_i \phi)^*  
\right) \ . 
\eqno(2.5)
$$

Eq.(2.2) is the gauged non-linear Schr\"odinger
equation, eq.(2.3) is Amp\`ere's law, a two-dimensional version of
$\nabla \times {\bf B} = {\bf J}$, and eq.(2.4) is a constraint.
There are
really two contributions to the total current, namely, the supercurrent
$J_i^S$ and the normal 
current $J^N_i$ which is present only in the cores of the vortices.
This being a non-dissipative model, there is no Ohmic
conductivity.  However, there is a Hall conductivity $2\mu$, and hence
$J^N_i = 2\mu \epsilon_{ij} E_j$. The effect of a Hall conductivity on
vortex motion was previously considered by Dorsey [16], and
equations (2.2)--(2.4) are just a special case of those analysed by Dorsey.

It is possible to have a constant, asymptotic value for the supercurrent
$J^S_i$.  This asymptotic current is called the transport current
$J_i^T$.  An everywhere constant current arises if $\phi = \exp \ i({\bf k}
\cdot {\bf x} - |{\bf k}|^2t/2\gamma)$, 
for example, and then ${\bf J}^T = {\bf k}$.
Such a constant current in a thin film produces a magnetic field above
and below the film (${1 \over 2}{\bf k} \times \hat{{\bf z}}$ above and
$-{1 \over 2}{\bf k} \times \hat{{\bf z}}$ below) but no field in the
film, and the three-dimensional Amp\`ere's law is thereby satisfied.  The
version of Amp\`ere's law (2.3) leaves out $\partial/\partial x_3$
terms.  To correct for this and avoid a linear growth of $B$ in the
plane, due to the constant current, it is necessary to subtract 
off ${\bf J}^T$ from ${\bf J}^S$, as in (2.3).  This also explains 
the need for the $-a_i J_i^T$ term in the Lagrangian density.  

The non-linear Schr\"odinger equation and its complex conjugate
imply the supercurrent conservation law
$$
\partial_i J^S_i + {\partial\over \partial t} (\gamma |\phi |^2 ) = 0
\ .
\eqno(2.6)
$$
In addition, there is Faraday's law
$$
\partial_1 E_2 - \partial_2 E_1 + {\partial B \over \partial t} = 0
\eqno(2.7)
$$
which is an immediate consequence of the definitions of $E_i$ and $B$
in terms of the gauge potential.  (The equation $\nabla \cdot
{\bf B} = 0 $
is trivially satisfied in the two-dimensional geometry.)  Taking the
divergence of (2.3), and combining (2.6) and (2.7), one finds
$$
{\partial\over \partial t} (2 \mu B + \gamma |\phi |^2) = 0 \ .
\eqno(2.8)
$$
To avoid explicitly breaking translational invariance,
$2\mu B + \gamma |\phi |^2$ must be a constant, independent of position; to
admit the condensate $|\phi | = 1$ with no magnetic field, this
constant must be $\gamma$.  Hence $2\mu B = \gamma (1- |\phi |^2)$, which
agrees with (2.4).

An interpretation of eq.(2.4) is that the total electric charge density
is zero. The charge density due to the
condensate is $- |\phi |^2$, and the charge density of the background positive
ions is $1$.  Where the condensate is absent, the background positive charge
is neutralised by decoherent, normal electrons; the normal
charge density is $- {2\mu \over \gamma} B$.
However, this is only an approximation, and the electric 
charge density is not exactly zero.
Note that there is no equation for $\partial_i E_i$ among our field
equations; this is because of the absence of ${1\over 2} E_i E_i$ in
the Lagrangian.  One may evaluate the charge density $\rho$
using $\partial_i E_i = \rho$; it is expected to be very small.  
Since Amp\`ere's
law requires that $\partial_i J_i^{\rm total}= 0$, one cannot use the
current conservation equation of electrodynamics to deduce anything
about $\rho$.  Physically, there will be small electric
charges generated by moving vortices, and $\partial_i J_i^{\rm
total}$ will not be exactly zero, but this can be ignored in the
non-relativistic limit.

A remarkable property of the system of equations (2.2)--(2.4), together
with (2.6) and (2.7), is that they are exactly Galilean invariant.  It
was stressed by Aitchison et al. that the equation for the scalar field
should be Galilean invariant [10].  Our additional equations define a
non-relativistic limit of Maxwell's equations with the same
invariance.  In fact, we have a ``magnetic version'' of Galilean
invariance, with substantial currents and negligible electric charge
density.  For an illuminating discussion of Galilean invariant limits
of electromagnetism, see ref. [17].

The basic Galilean transformation is a boost by a velocity ${\bf v}$.
Gauge invariant scalar quantities transform as $f \rightarrow f'$
where  $f' ({\bf x}, t) = f ({\bf x} - {\bf v}t, t)$.  Let us denote
${\bf x}' = {\bf x} - {\bf v}t$.  The field transformations are
$$
\eqalignno{
\phi ' ({\bf x},t) &= \phi ({\bf x}', t) e^{i \gamma ({\bf v}\cdot
{\bf x} - {1\over 2} |{\bf v}|^2 t)}&(2.9) \cr
a_i' ({\bf x}, t) &= a_i ({\bf x}', t) &(2.10) \cr 
a_t' ({\bf x}, t) &= a_t ({\bf x}', t) - v_i a_i ({\bf
x}', t) \ . &(2.11) \cr
}
$$
It is straightforward to verify that eq.(2.2) is invariant under these
transformations, in the same way that the Schr\"odinger equation in
quantum mechanics is Galilean invariant when there is no
potential.  The magnetic and electric fields transform to
$$
\eqalignno{
B' ({\bf x}, t) &= B({\bf x}', t) &(2.12) \cr
E'_i ({\bf x}, t) &= E_i ({\bf x}', t) - \epsilon_{ij} v_j B({\bf
x}', t) &(2.13) \ , \cr
}
$$
and the supercurrent becomes
$$
{\bf J}'^{S} ({\bf x}, t) = {\bf J}^S ({\bf x}', t) + \gamma {\bf v}
| \phi ({\bf x}', t) |^2 \ . \eqno(2.14)
$$
The transport current has therefore also transformed to
$$
{\bf J}'^{T} = {\bf J}^{T} + \gamma{\bf v} \ .
\eqno(2.15)
$$
Combining eqs.(2.12)--(2.15), we find that eq.(2.3) is Galilean invariant
provided the constraint (2.4) is satisfied.  Finally, it is easily
checked that the constraint (2.4) itself remains unchanged by a
Galilean transformation.

The physical interpretation of the Galilean invariance is as follows.
Given any solution of eqs.(2.2)--(2.4) in the absence of a
transport current, the effect of a transport current ${\bf J}^T$ is 
simply to boost
the solution so that it drifts along with the current at a velocity
${\bf v} = {1\over \gamma} {\bf J}^T$.

Since we now understand the effect of a transport current, let us
from now on assume there isn't one, and that the fields have finite
energy, with all currents and electromagnetic fields localized in a
finite region of space.

{\bf 3. Vortices}

In the Ginzburg--Landau theory, fields which are smooth and of finite
energy have the asymptotic behaviour $|\phi | \rightarrow 1$ and
$D_i \phi \rightarrow 0$ at spatial infinity. Such fields are
classified by their integer winding number. When the winding number is
$n$, the phase of $\phi$
increases by $2\pi n$ anticlockwise around the circle at infinity. 
The vanishing of $D_i \phi$ at
infinity implies that the gauge potential also carries information
about the winding number, from which follows the magnetic
flux quantization
$$
\int B \ d^2x = 2\pi n \ .
\eqno(3.1)
$$
If the zeros of $\phi$ are isolated, then the net number of zeros,
counted with their multiplicity, is also $n$. A zero
of multiplicity $1$ may be identified as a magnetic flux vortex, and a zero of 
multiplicity $-1$ as an
antivortex, so the winding number is also the net vortex number.

The static Ginzburg--Landau equations
$$
\eqalignno { 
D_i D_i \phi + {\lambda \over 2} (&1 - |\phi |^2 ) \phi = 0 &(3.2) \cr
\epsilon_{ij} \partial_j B &= J^S_i &(3.3) \cr
}
$$
have a vortex solution with unit winding number of the
form (in polar coordinates)
$$
\eqalign{
\phi &= k(r) e^{i\theta} \cr
a_1 = -{g(r) \over r}\sin \theta \ &, \quad  a_2 = {g(r) \over r}
\cos \theta \ , \cr
}
\eqno(3.4) 
$$
where $k \rightarrow 1$ and $g \rightarrow
1$ exponentially fast as $r \rightarrow \infty$, and where 
$k$ and $g$ both vanish at $r=0$.  The precise form of
$k$ and $g$ must be determined numerically.  More generally, for
winding number $n > 1$, there are
multivortex static solutions of a similar form, but with $\phi = k(r)
e^{in\theta }$, and where $k \rightarrow 1$ and $g \rightarrow
n$ as $r \rightarrow \infty$.
If $\lambda > 1$, 
vortices repel, so the circularly symmetric multivortex solution is
unstable to break-up into individual vortices.  

In the special case $\lambda = 1$, there are no static
forces between vortices.  As Bogomolny showed [18], in this case eqs.(3.2)
and (3.3) are satisfied, and the energy minimised, provided the first
order Bogomolny equations
$$
\eqalignno{
D_1 \phi &+ i D_2 \phi = 0 &(3.5) \cr
B &= {1\over 2} (1 - |\phi |^2 ) &(3.6) \cr
}
$$
are satisfied. Eq.(3.5) only permits $\phi$ to have zeros of positive
multiplicity, so solutions consist only of vortices. (To get
antivortices, the sign in (3.5) and the sign of $B$ should be reversed.)  
The Bogomolny equations have not only circularly symmetric solutions,
but, for any $n \geq 1$, solutions with vortices located at $n$
arbitrary points in the
plane.  More precisely, Taubes proved that, modulo gauge
transformations, there is a unique finite energy solution of
eqs.(3.5) and (3.6) with $\phi$ having zeros at any $n$ prescribed points
$ \{ {\bf x}^r : 1 \leq r \leq n \}$, some of which may
coincide [19].  Each such solution has winding $n$, total flux $2\pi n$,
and energy $\pi n$.

The parameter space of such solutions, which is called the 
$n$-vortex moduli space
$M^n$, is a smooth manifold of dimension
$2n$.  To see
this, identify ${\Bbb R}^2$ and ${\Bbb  C}$, and regard the zeros of
$\phi$ as the complex numbers $\{z_r : 1 \leq r \leq n
\}$. Geometrically, $M^n$ is
the manifold ${\Bbb C}^n /\Sigma^n$ where $\Sigma^n$ is the
permutation group acting on the $n$ zeros of $\phi$.  ${\Bbb C}^n/
\Sigma^n$ is actually smooth despite the apparent conical
singularities where two or more zeros of $\phi$ coincide. The
zeros uniquely define a polynomial $P(z)
= z^n + a_1 z^{n-1} + \dots + a_n$ with precisely these zeros.  
The coefficients $\{ a_1, \dots ,a_n \}$, which are symmetric
polynomials in the zeros, 
are arbitrary complex numbers, so, as a manifold, $M^n$ is also the space of
monic (leading coefficient $=1$) polynomials of degree $n$, and
this is simply ${\Bbb C}^n$. The coefficients $\{ a_1, \dots ,a_n \}$
rather than the unordered zeros $\{ z_1, \dots ,z_n \}$
are the ``good''  coordinates on ${\Bbb C}^n / \Sigma^n$. 
 
The equations (2.2)--(2.4) that we are interested in are not
simply the static Ginzburg--Landau equations. However, if $\lambda =
1$ and $\mu = \gamma$, any solution of the Bogomolny equations 
(3.5)--(3.6) is also a
static solution of eqs.(2.2)--(2.4).  The constraint (2.4) is one of the
Bogomolny equations, so it is satisfied.  
$D_t \phi$ and $E_i$ can be consistently set to zero, with $a_t = 0$.

One might seek static solutions of eqs.(2.2)--(2.4) for $\lambda
\neq 1$.  Presumably such solutions exist, if $a_t$ is allowed to be
non-zero.  They represent stationary points of the Ginzburg-Landau
energy, subject to the constraint (2.4), with $a_t$ a Lagrange
multiplier field.  Solutions of the Bogomolny equations
will come close to being solutions if $\lambda \simeq
1$ and $\mu \simeq \gamma$.  As in the earlier
case, with  $\lambda \neq 1$, only circularly symmetric 
static solutions are expected, and they may again be unstable.

The constraint (2.4) is very far from being satisfied by the static
solutions of the Ginzburg--Landau equations in the extreme Type II
regime ($ \lambda
\gg 1$).  It is therefore unclear whether our model is of any
relevance to vortex dynamics in this regime.  Instead, we shall
consider the case of $\lambda \simeq 1$, which is realised by
niobium and vanadium in certain temperature ranges.

{\bf 4. A Reduced Lagrangian}

The rest of this paper is devoted to constructing a reduced Lagrangian
for $n$-vortex dynamics, assuming $\lambda \simeq 1$.  We need to
assume that $\mu = \gamma$; this is essential for simplifying the
kinetic energy. The method is
similar to that used to study $n$-vortex dynamics in the Abelian Higgs
model, both at $\lambda = 1$ and when $\lambda \simeq 1$.  Let us consider
fields which at each instant are
static solutions of the Bogomolny equations, but where the 
moduli, that is, the vortex positions ${\bf x}^r$
(or better, the coefficients of the polynomial $P(z)$)
are time-dependent.  These moduli will vary slowly if $\lambda$ is
close to $1$.  The fields are
inserted in the Lagrangian (2.1), and the integrals carried out, where
possible. The result is a reduced Lagrangian for motion on the moduli
space $M^n$. The reduced system is an approximation to the true field
dynamics, but we shall not try to estimate the errors involved.  The
use of solutions of the Bogomolny equations is possibly better
justified here than in the context of the Abelian Higgs model, because
the constraint (2.4), which is one of the Bogomolny equations, must be
satisfied.

Let us denote by $\{ X^{\alpha} : 1 \leq \alpha \leq 2n \}$ some
general coordinates on the moduli space $M^n$, for example,
the components of the vortex positions.  
The field Lagrangian has a
kinetic term which is first order in time derivatives, and a potential
term.  The reduced system is therefore expected to have a Lagrangian of the
form 
$$
L = {\cal A}_{\alpha} ({\bf X}) \dot X^{\alpha} - V({\bf X}) \ .
\eqno(4.1)
$$
$L$ may have additional terms which are total time derivatives, but
these do not affect the dynamics. ${\cal A}_{\alpha}$  has the 
interpretation of a gauge potential or connection on $M^n$, 
and it is somewhat arbitrary since replacing it by ${\cal A}_{\alpha}'
= {\cal A}_{\alpha} 
+ \partial_{\alpha} {\mit \Lambda}$ (a gauge transformation) changes $L$ by
the total time derivative ${d {\mit \Lambda} \over d t}$.  From $L$ we find the
equations of motion
$$
{\cal B}_{\alpha \beta} \dot{X}^{\alpha} + \partial_{\beta} V = 0
\eqno(4.2)
$$
where ${\cal B}_{\alpha \beta} = \partial_{\alpha} {\cal A}_{\beta} - 
\partial_{\beta} {\cal A}_{\alpha} $ is the curvature of the
connection ${\cal A}$.
The motion according to (4.2) is non-dissipative, with $V$ constant
along any solution path $X^{\alpha} (t)$.

The potential energy in the reduced model is the Ginzburg-Landau
potential energy, but because the fields satisfy the Bogomolny
equations, this simplifies to [18]
$$
V = n\pi + {{\lambda - 1} \over 8}\int(1 - |\phi |^2)^2 \ d^2x \ .
\eqno(4.3)
$$
The constant $n\pi$ does not affect the dynamics. The
integral without the factor ${1 \over 8}(\lambda -1)$ is positive, and 
is some function of the relative positions of the
vortices, invariant under a rigid rotation. The detailed form of the
integral is not known, for general $n$, but it is expected to be
minimal when the vortices are well separated, and maximal when the
vortices are coincident. The integral has been computed in the case of
two vortices, by Shah [13]. It increases monotonically as the vortex
separation decreases.

We can proceed much further with the calculation of the kinetic energy
in the reduced model. Let us start with the kinetic terms of the field
Lagrangian, with $\gamma = \mu$
$$
T=\gamma\int\Bigl( {i \over 2}(\phi^*\dot\phi - \phi\dot\phi^*) 
+ |\phi|^2 a_t - a_t + B a_t + 
(\partial_2 a_t - \dot a_2) a_1 - 
(\partial_1 a_t - \dot a_1) a_2 \Bigr) \ d^2x \ . 
\eqno(4.4)
$$
It helps to make some assumptions about the
asymptotic gauge. Recall that vortices are exponentially
localized. Gauge invariant quantities $B$, $|D_i\phi |$ and $|\phi |$
approach their asymptotic values exponentially fast. Suppose
that the solutions of the Bogomolny equations are in the gauge where,
for large $r$,
they are of the form $\phi = e^{in\theta}, \ a_1 =
-{n \over r}\sin\theta, \ a_2 = {n \over r}\cos\theta$, with 
at most exponentially small corrections. Since $E_i$ and 
$D_t \phi$ are exponentially small 
asymptotically, in this gauge $a_t$ is exponentially small too.

Using the appropriate version of Stokes' theorem, the terms involving
derivatives of $a_t$ can now be removed, and $T$ expressed as
$$
T=\gamma\int\Bigl( {i \over 2}(\phi^*\dot\phi - \phi\dot\phi^*)+a_2\dot a_1
- a_1\dot a_2\Bigr) \ d^2x
+ \gamma\int (|\phi |^2 - 1 + 2B)a_t \ d^2x \ .
\eqno(4.5)
$$
The boundary terms vanish as a result of the gauge fixing,
and the Bogomolny equation (3.6) (or the constraint (2.4)) implies
that the second integral vanishes. Now define $a = a_1 + ia_2$. 
Then $T$ reduces to the compact expression
$$
T = -\gamma\int {\rm Im} (\phi^*\dot\phi + a^*\dot a) \ d^2x \ .
\eqno(4.6)
$$

Let us next express the scalar field as
$$
\phi = e^{{1 \over 2}h + i\chi}
\eqno(4.7)
$$
where $h$ and $\chi$ are real. $\phi$ is a smooth function, but the
gauge is not fixed except asymptotically, so there is considerable
freedom in the choice of $\chi$. Because of the presence of vortices,
$\chi$ is multivalued, increasing by an integer multiple of 
$2\pi$ along an anticlockwise
loop around any zero of $\phi$. However, in a given gauge, the
gradient and time derivative of $\chi$ are well-defined.

Away from the zeros of $\phi$, the Bogomolny equation (3.5) implies
that
$$
a = {1 \over 2} \partial_2h + \partial_1\chi + 
i(-{1 \over 2} \partial_1h + \partial_2\chi) 
\eqno(4.8)
$$
and hence
$$
\dot a = {1 \over 2} \partial_2\dot h + \partial_1\dot\chi + 
i(-{1 \over 2} \partial_1\dot h + \partial_2\dot\chi) \ . 
\eqno(4.9)
$$
Also
$$
\dot\phi = ({1 \over 2}\dot h + i\dot\chi)e^{{1 \over 2}h + i\chi } \ .
\eqno(4.10)
$$
Therefore
$$
T = -\gamma\int \Bigl( e^h\dot\chi + 
({1 \over 2} \partial_2h + \partial_1\chi)
(-{1 \over 2} \partial_1\dot h + \partial_2\dot\chi) 
- (-{1 \over 2} \partial_1h + \partial_2\chi)
({1 \over 2} \partial_2\dot h + \partial_1\dot\chi) \Bigr) 
\ d^2x \ .
\eqno(4.11)
$$
It is easy to verify that (4.11) can be reexpressed as
$$
T = -\gamma\int \Bigl( (e^h - \nabla^2h)\dot\chi - {1 \over
2}{\partial \over {\partial t}}(\nabla h\cdot\nabla\chi) + \partial_2 f_1 -
\partial_1 f_2 \Bigr) \ d^2x 
\eqno(4.12)
$$
where
$$
f_1 = (\partial_2h + \partial_1\chi)\dot\chi 
+ {1 \over 4}\partial_1h\dot h \ , \quad \quad
f_2 = (-\partial_1h + \partial_2\chi)\dot\chi 
+ {1 \over 4}\partial_2h\dot h \ .
\eqno(4.13)
$$
Since, from (4.8), $B = -{1 \over 2}\nabla^2h$, the Bogomolny equation (3.6)
reduces to
$$
\nabla^2h - e^h + 1 = 0 \ ,
\eqno(4.14)
$$
so $T$ simplifies to
$$
T = -\gamma\int \Bigl(\dot\chi - {1 \over
2}{\partial \over {\partial t}}(\nabla h\cdot\nabla\chi) + \partial_2 f_1 -
\partial_1 f_2 \Bigr) \ d^2x \ .
\eqno(4.15)
$$

To progress, we need to specify carefully the region of
integration. It will simplify our calculations to assume that at any
time $t$, the vortex
positions ${\bf x}^r (t)$ are distinct, so $\phi$
has a simple zero at each of these $n$ points.  The case of two or more of
these points becoming coincident can be treated by taking a limit.
The fields $\phi$ and $a$ are smooth functions of
space and time, so the original expression (4.4) for $T$ has no
singularities. However, $h$ and $\chi$ are ill-defined at the
(moving) vortex locations, and $\nabla h$, $\nabla \chi$, $\dot h$ and
$\dot\chi$ all diverge as a vortex is approached. Let us therefore
define the region of integration $\Sigma$ to be that obtained from 
$\Bbb R^2$ by removing $n$ discs of small radius
$\epsilon$, centred at the vortex locations ${\bf x}^r(t)$. This
produces an error of order $\epsilon^2$ in $T$. At the end of the
calculation, let $\epsilon \rightarrow 0$. During the
calculation we can neglect any terms in $T$ (but not in the component
fields) that are $O(\epsilon)$ or smaller.

Let $C^r(\epsilon)$
denote the circular boundary of the $r$th disc. The terms involving $f_1$
and $f_2$ reduce to line integrals along
$C^r(\epsilon)$. There is no contribution from infinity as $f_1$ and $f_2$
decay exponentially fast. The integral of 
${\partial \over {\partial t}}(\nabla h\cdot\nabla\chi)$
can be expressed in terms of a total time derivative. We have
$$
{d \over dt} \int_\Sigma \nabla h\cdot\nabla\chi \ d^2x =
\int_\Sigma {\partial \over {\partial t}} (\nabla h\cdot\nabla\chi) \
d^2x - \sum_{r=1}^n
\int_{C^r(\epsilon)}(\nabla h\cdot\nabla\chi) \ \dot {\bf x}^r \times {\bf
dl} \ ,
\eqno(4.16)
$$
taking into account that the boundary $C^r(\epsilon)$ is
moving with velocity $\dot {\bf x}^r$. The line element ${\bf dl}$ is that
along $C^r(\epsilon)$, and $\dot {\bf x}^r \times {\bf dl}$ is a
scalar (in two dimensions).
If we drop the total time derivative from $T$, which does not affect
the dynamics, the integral of 
${\partial \over {\partial t}}(\nabla h\cdot\nabla\chi)$ is simply 
replaced by integrals over the circles $C^r(\epsilon)$.

Combining these observations we find
$$
T = -\gamma\int_{\Sigma} \dot\chi \ d^2x \ + {\gamma \over 2}\sum_{r=1}^n
\int_{C^r(\epsilon)}(\nabla h\cdot\nabla\chi) \ \dot {\bf x}^r \times {\bf
dl} \ - \gamma\sum_{r=1}^n \int_{C^r(\epsilon)} {\bf f}\cdot {\bf dl} \ .
\eqno(4.17)
$$
To calculate the integrals along $C^r(\epsilon)$ we 
need the expansions of $h$ and $\chi$ around the
vortex centre ${\bf x}^r$. Since $\phi$ is linear in the 
neighbourhood of ${\bf x}^r$, to first
approximation, $h$ has the
expansion [11]
$$
h = \log |{\bf x} - {\bf x}^r(t)|^2 + \alpha^r + \beta_1^r(x_1 -
x_1^r(t)) + \beta_2^r(x_2 - x_2^r(t)) + \dots
\eqno(4.18)
$$
where $\alpha^r$, $\beta_1^r$ and $\beta_2^r$ depend on the locations
of the other vortices, and vary smoothly with
time as the vortices move. The contours of $|\phi |$ are 
approximately circles near ${\bf x}^r$,
and ($\beta_1^r , \beta_2^r$) measures the extent to which the centres
of these circles differ from ${\bf x}^r$ as $|\phi |$ increases,
because of the other vortices.

From (4.18), we find that near ${\bf x}^r$
$$
\nabla h = \left( {2(x_1 - x_1^r) \over |{\bf x} - {\bf x}^r |^2} +
\beta_1^r \ , \ {2(x_2 - x_2^r) \over |{\bf x} - {\bf x}^r |^2} +
\beta_2^r \right)
\eqno(4.19)
$$
and
$$
\dot h = - \left( {2(x_1 - x_1^r) \over |{\bf x} - {\bf x}^r |^2} +
\beta_1^r \ \right)\dot x_1^r \ - \ 
\left( {2(x_2 - x_2^r) \over |{\bf x} - {\bf x}^r |^2} +
\beta_2^r \ \right)\dot x_2^r \ + \dot\alpha^r + \dot\beta_1^r(x_1 -
x_1^r) + \dot\beta_2^r(x_2 - x_2^r)
\eqno(4.20)
$$
with higher order corrections that can be neglected.
Let us introduce a polar angle $\theta^r$ relative to the (moving)
vortex at ${\bf x}^r$, with $\theta^r = 0$ in the positive $x_1$ 
direction. Then on $C^r(\epsilon)$
$$
\nabla h = \left( {2\cos \theta^r \over \epsilon} +
\beta_1^r \ , \ {2\sin\theta^r \over \epsilon} +
\beta_2^r \right)
\eqno(4.21)
$$
and
$$
\dot h = - \left( {2\cos\theta^r \over \epsilon} +
\beta_1^r \ \right)\dot x_1^r \ - \ 
\left( {2\sin\theta^r \over \epsilon} +
\beta_2^r \ \right)\dot x_2^r \ + \dot\alpha^r 
+ \epsilon\left(\dot\beta_1^r \cos\theta^r + \dot\beta_2^r
\sin\theta^r \right) \ .
\eqno(4.22)
$$

Next, suppose that the gauge has been chosen so that on and inside
the circle $C^r(\epsilon)$, the phase $\chi$ of the field $\phi$ is exactly
linearly dependent on the polar angle $\theta^r$, that is
$$
\chi = \theta^r + \psi^r
\eqno(4.23)
$$
where $\psi^r$, which will be referred to as the orientation of vortex $r$, 
depends only on time. 
Then, on $C^r(\epsilon)$
$$
\nabla \chi = \left( - {\sin\theta^r \over \epsilon} \ , \ 
{\cos\theta^r \over \epsilon}\right)
\eqno(4.24)
$$
and
$$
\dot\chi = {1 \over \epsilon} (\dot x_1^r \sin\theta^r - \dot x_2^r
\cos\theta^r ) + \dot\psi^r \ .
\eqno(4.25)
$$
We shall see that it is not
possible globally to set $\psi^r = 0$, although it would be on a short
time interval.

We can now evaluate the line integrals along $C^r(\epsilon)$ in the
kinetic energy expression (4.17). Note that ${\bf dl} = \epsilon
(-\sin\theta^r , \cos\theta^r )\ d\theta^r$, and $\dot {\bf x}^r \times {\bf
dl} \ = \epsilon (\dot x_1^r \cos\theta^r + \dot x_2^r \sin\theta^r )
\ d\theta^r $.
Therefore
$$\eqalign{
\int_{C^r(\epsilon)}(\nabla h\cdot\nabla\chi)\dot {\bf x}^r \times {\bf
dl} \ &= \int_0^{2\pi}\left(-\beta_1^r \sin\theta^r + \beta_2^r
\cos\theta^r \right)\left(\dot x_1^r \cos\theta^r   + \dot x_2^r \sin\theta^r
\right) d\theta^r \cr &= \pi (\beta_2^r \dot x_1^r - \beta_1^r \dot
x_2^r) \ , \cr}
\eqno(4.26)
$$
and
$$\eqalign{
\int_{C^r(\epsilon)} {\bf f}\cdot{\bf dl} &= - \int_0^{2\pi}\biggl( 1 +
\epsilon (\beta_2^r \sin\theta^r + \beta_1^r
\cos\theta^r ) \biggr)\biggl( {1 \over \epsilon}(\dot x_1^r
\sin\theta^r   -  \dot x_2^r \cos\theta^r ) + \dot\psi^r \biggr) d\theta^r
\cr 
&\quad \quad \quad 
-{1 \over 2}\int_0^{2\pi}\left(-\beta_1^r \sin\theta^r + \beta_2^r
\cos\theta^r \right)\left(\dot x_1^r \cos\theta^r  + \dot x_2^r \sin\theta^r
\right) d\theta^r \cr
&= -2\pi\dot\psi^r 
- {3\pi \over 2}(\beta_2^r \dot x_1^r - \beta_1^r \dot x_2^r) \ ,\cr}
\eqno(4.27)
$$
so
$$
T = -\gamma\int_{\Sigma} \dot\chi \ d^2x \ 
+ 2\pi\gamma\sum_{r=1}^n \bigl( \dot\psi^r \ 
+ (\beta_2^r \dot x_1^r - \beta_1^r \dot x_2^r) \bigr)
\ .
\eqno(4.28)
$$

We still need to consider the integral of $\dot\chi$. This is not the time
derivative of the integral of $\chi$, since $\chi$ is multivalued and
its integral over the plane ill-defined. However, a gauge
transformation replaces $\chi$ by $\chi + \tilde\chi$ where
$\tilde\chi$ is single-valued, and the integral of $\dot\chi$ changes
by the time derivative of the integral of $\tilde\chi$. So, up to a
total time derivative, the integral of $\dot\chi$ is gauge
invariant. It is convenient, in this integral, to extend the region of
integration back to the whole plane. $\dot\chi$ is 
$O({1 \over \epsilon})$ 
near the vortices, so the contribution to its integral from the
discs of radius $\epsilon$ is $O(\epsilon )$ and can be neglected.

The integral cannot be evaluated directly. Instead, consider its
integral over a finite time interval
$$
\int_{t_0}^{t_1} \int_{\Bbb R^2} \dot\chi \ d^2x \ dt \ .
\eqno(4.29)
$$
Suppose that the initial and final configurations are the same, that
is, the vortex locations are the same, and the fields are in the same
gauge. Consider first the simple case where just one vortex moves
anticlockwise around a loop which does not enclose any other
vortices. At a point outside the loop, $\chi$ varies but there is no
net change in $\chi$ between the initial and final time. (At infinity
this is true, and by continuity this result extends to any point
outside the loop.) At a point inside the loop, on the other hand,
$\chi$ increases by $2\pi$. This can be verified by deforming the
motion of the vortex around the point into an anticlockwise motion of
the point around the vortex. The integral (4.29) is therefore $2\pi$ 
times the area of the loop. For a more general closed vortex
trajectory, possibly with self-crossings, the integral is $2\pi$ times
the signed area enclosed.

The result generalizes. Even if the loop encloses other non-moving
vortices, the integral is $2\pi$ times the area of the loop. If all
$n$ vortices move, the integral is $2\pi$ times the sum of the
areas enclosed by the $n$ vortex loops. Finally, we must
allow for the vortices to exchange locations. Suppose two vortices
move anticlockwise along half-loops, such that they exchange
places. Then the integral is $2\pi$ times the area
enclosed by the loop. All these results can be checked by using a
specific model for $\chi$, for example the phase of the complex
polynomial $P(z) = \prod_{r=1}^n (z - z_r(t))$, where $z$ and $z_r(t)$
are the complex numbers representing a general point in the plane and
the trajectory of the $r$th vortex. Then it is easy to calculate the
change in $\chi$ at $z$ due to the vortex motion. Unfortunately, this
model doesn't quite satisfy our requirements on the phase at infinity
or on the circles $C^r(\epsilon)$,
but this can be dealt with easily.

Now observe that the correct value for the integral of $\dot\chi$
over space and time, as discussed above, can be obtained from the
local expression
$$
\pi\sum_{r=1}^n (-x_2^r\dot x_1^r + x_1^r\dot x_2^r) \ ,
\eqno(4.30)
$$
whose integral over time gives again $2\pi$ times the sum of the
(signed) areas enclosed by the vortex trajectories. So the expression
(4.30) is equal to the integral of $\dot \chi$ over the plane, up to
an ignorable time derivative. Therefore $T$ can
be rewritten as
$$
T = 2\pi\gamma\sum_{r=1}^n \left( \dot\psi^r + (\beta_2^r + {1 \over
2}x_2^r)\dot x_1^r - (\beta_1^r + {1 \over 2}x_1^r)\dot x_2^r \right)
\ ,
\eqno(4.31)
$$
and this is our final expression for $T$. The term (4.30) has occurred
before, in the context of ungauged vortices [20], although the neat
calculation in [20] involves the manipulation of a divergent integral.
For well separated vortices, the ${\bf x}^r$ terms in (4.31) dominate,
since $\beta_1^r$ and $\beta_2^r$ are exponentially small. The
coefficients $\beta_1^r$ and $\beta_2^r$ have a significant effect 
when the vortex cores are overlapping. Their appearance is not
surprising as they also appear in the expression for the
Riemannian metric on $M^n$, and Samols has determined some of their
properties which will not all be used here, but which could be useful [11]. 

The $\dot\psi^r$ terms are topological. Along a closed path in $M^n$,
with the initial and final configurations identical, the initial and
final values of the orientations $\psi^r$ of the vortices are
geometrically the same, but they may have been permuted, and they may
also have changed by integer multiples of $2\pi$. The motion of the
vortices relative to each other defines a braid, and the integral over
time of
$\sum_{r=1}^n \dot\psi^r$ can depend only on this braid. In fact, each
positive generator of the braid group contributes $2\pi$ to the
integral, as we shall see below.

There is an important consistency check on the expression (4.31) for
$T$. Note that, as anticipated, $T$ defines a connection on the moduli
space $M^n$. The integral of this connection around any closed loop on
$M^n$ is called the holonomy around that loop. It can be calculated by
integrating $T$ over time, for motion (at any speed) around the
loop. Since $M^n$ has no singularities, the holonomy should vanish as
a loop contracts to a point. In particular, consider the closed loop
on $M^2$ where two vortices
exchange places by moving anticlockwise along semi-circular
trajectories. 
This motion is a generator of the braid group. The
holonomy should become zero as the radius shrinks to zero. Let the
vortex trajectories be ${\bf x}^1(t) = R(\cos t, \sin t)$ and 
${\bf x}^2(t) = -R(\cos t, \sin t)$, with $0 \le t \le \pi$. The
complex polynomial representation $P(z) = (z - z_1(t))(z - z_2(t))$
shows that the orientations of the two vortices are $\psi^1(t) =t$ and
$\psi^2(t) = \pi + t$, so the integral of $\dot \psi^1 + \dot \psi^2$
is $2\pi$. Also, by
circular and reflection symmetry, $(\beta_1^1 , \beta_2^1)$ is of the
form $\beta(R)(\cos t,\sin t)$ and $(\beta_1^2 , \beta_2^2) =
-(\beta_1^1 , \beta_2^1)$. A simple calculation shows that the total
holonomy is
$$
-2\pi\gamma(\pi R^2 - 2\pi + 2\pi\beta(R)R) \ .
\eqno(4.32)
$$

The coefficient $\beta(R)$ is defined as that occurring in the
expansion of $h$ around ${\bf x}^1 = (R,0)$ when the vortices are at
$(R,0)$ and $(-R,0)$.  By reflection symmetry, this expansion has the form
$$
h = \log((x_1 - R)^2 + (x_2)^2) + \alpha(R) + \beta(R)(x_1 -R) + \dots \
.
\eqno(4.33)
$$
For large $R$, the vortex at $(-R,0)$ has an exponentially small
effect, so $\beta(R)$ is exponentially small and the holonomy is
proportional to $\pi R^2 -2\pi$, the $2\pi$ being a topological
correction to the area of the circle enclosed by the vortex trajectories. When
$R$ is small, $h$ can be estimated simply from the leading logarithmic
terms due to both vortices
$$
h = \log((x_1 - R)^2 + (x_2)^2) + \log((x_1 + R)^2 + (x_2)^2) + \dots \ .
\eqno(4.34)
$$
Expanding about $(R,0)$, this becomes
$$
h = \log((x_1 - R)^2 + (x_2)^2) + \log 4R^2 + {1 \over R}(x_1 - R) + \dots \ ,
\eqno(4.35)
$$
so $\beta(R) = {1 \over R}$ for small $R$. This singular behaviour is
just what is needed for the holonomy to vanish as $R \rightarrow 0$.

Presumably, a more sophisticated version of this argument would
establish that the holonomy vanishes for any loop on $M^n$ as the loop
shrinks to a point.

{\bf 5. Vortex Motion}

The Lagrangian derived in section 4 for the motion of $n$ vortices is
$$
L = 2\pi\gamma\sum_{r=1}^n \left( \dot\psi^r + (\beta_2^r + {1 \over
2}x_2^r)\dot x_1^r - (\beta_1^r + {1 \over 2}x_1^r)\dot x_2^r \right) 
- V({\bf x}^1,{\bf x}^2,\dots ,{\bf x}^n) \ ,
\eqno(5.1)
$$
where $V$ is the integral expression (4.3). The term $\dot \psi^r$ is
locally a time derivative, and so has no effect on the vortex motion,
although it has a topological significance as we have seen. A rather
similar Lagrangian has been obtained for well separated vortices by
Dziarmaga [21], using Berry phase methods.

The general form of the equation of motion for vortex $r$ is
$$
{d \over dt}\left(\partial L \over \partial \dot x_i^r\right) - 
{\partial L \over \partial x_i^r} = 0 \ .
\eqno(5.2)
$$
For $L$ as above, this becomes
$$\eqalign{
2\pi\gamma\left[\dot x_1^r + \left({\partial \beta_1^r \over \partial x_1^s} + 
{\partial \beta_2^s \over \partial x_2^r}\right)\dot x_1^s + 
\left({\partial \beta_1^r \over \partial x_2^s} - 
{\partial \beta_1^s \over \partial x_2^r}\right)\dot x_2^s\right] - {\partial V
\over \partial x_2^r} &= 0 \cr
2\pi\gamma\left[\dot x_2^r + \left({\partial \beta_2^r \over \partial x_2^s} + 
{\partial \beta_1^s \over \partial x_1^r}\right)\dot x_2^s + 
\left({\partial \beta_2^r \over \partial x_1^s} - 
{\partial \beta_2^s \over \partial x_1^r}\right)\dot x_1^s\right] + {\partial V
\over \partial x_1^r} &= 0 \ , \cr}\eqno(5.3)
$$
with summation over $s$ implied.
It is straightforward to check that $V$ is constant along a
trajectory. This implies that if $n$ vortices are initially
well separated, they remain so, and similarly if they are initially all close
together.

Symmetries imply further conservation laws. Consider the Lagrangian
for motion on $M^n$ in the general form (4.1). If $\xi^{\alpha}({\bf
X})$ is the vector field generating a symmetry (i.e. the Lie
derivatives ${\cal L}_{\xi}{\cal A}$ and ${\cal L}_{\xi}V$ vanish),
then ${\cal A}_{\alpha}({\bf X})\xi^{\alpha}({\bf X})$ is a constant
of the motion. The symmetries here are translations and rotations
in the plane. The associated conserved quantities that follow from the
specific Lagrangian (5.1) are, respectively,
$$
\sum_{r=1}^n (\beta_1^r + {1 \over 2}x_1^r) \ , \ \
\sum_{r=1}^n (\beta_2^r + {1 \over 2}x_2^r)
\eqno(5.4)
$$
and
$$
\sum_{r=1}^n \left( (\beta_1^r + {1 \over 2}x_1^r)x_1^r + 
(\beta_2^r + {1 \over 2}x_2^r)x_2^r \right) \ .
\eqno(5.5)
$$
Samols [11] has shown that for any locations of the vortices,
$$
\sum_{r=1}^n \beta_1^r = \sum_{r=1}^n \beta_2^r = 0 \ ,
\eqno(5.6)
$$
so the conservation of the quantities (5.4) implies that the centre of
the $n$-vortex system, ${1 \over n}\sum_{r=1}^n {\bf x}^r$, is fixed.

If there is just one vortex, it does not move. 
This is because $\beta_1$ and $\beta_2$
vanish, by circular symmetry or by (5.6), and $V$ is independent of the vortex
position. 

Two vortices move around each other in a circular motion,
rather like two fluid vortices of equal strength. To see this,
change coordinates, writing
$$\eqalign{
{\bf x}^1 &= {\bf X}^0 + R(\cos\theta ,\sin\theta) \cr
{\bf x}^2 &= {\bf X}^0 - R(\cos\theta ,\sin\theta) \ . \cr} \eqno(5.7)
$$
As was remarked in the previous section, $(\beta_1^1,\beta_2^1) 
= -(\beta_1^2,\beta_2^2) = \beta(R)(\cos\theta ,\sin\theta)$. 
The potential $V$ depends only on $R$.
In these coordinates, the Lagrangian simplifies to
$$
L = 2\pi\gamma\left(2\dot\theta + X_2^0\dot X_1^0 -  X_1^0\dot X_2^0 - R(R + 
2\beta(R))\dot\theta \right) - V(R) \ . 
\eqno(5.8)
$$
The term $2\dot\theta$ equals $\dot\psi^1 + \dot\psi^2$. It has been
included, although it does not affect the equations of
motion, which are
$$\eqalign{
\dot X_1^0 = \dot X_2^0 = 0 \ , \ \dot R &= 0 \cr
2\pi\gamma\left({d \over dR}(R^2 + 2R\beta(R)) \right) &\dot\theta = -{dV
\over dR} \ . \cr } 
\eqno(5.9)
$$
The centre ${\bf X}^0$ does not move, and the relative motion is at
constant angular velocity on a fixed circle.

At large separation, ${dV \over dR}$ and $\beta$ are exponentially small, so
$\dot\theta$ is exponentially small. The maximum angular velocity
occurs at a finite separation of order the vortex core size.
At small separation, ${dV \over
dR} \simeq O(R^3)$. This is because $V$ has a maximum at $R=0$, and
varies quadratically with the ``good'' radial coordinate on $M^2$,
which is $R^2$. The $1 \over R$ singularity in $\beta(R)$ does not
produce a singular coefficient of $\dot\theta$, so the angular
velocity vanishes as $R$ tends to zero. In the limit, the two vortices
are coincident and at rest. It would be a useful
consistency check to see if there is a corresponding
exact, static solution of the field equations.

Symmetry implies that if there are $p$ vortices at the vertices of a
regular $p$-gon, and $q$ vortices at the centre (possibly with $q=0$),
then the $p$-gon will rigidly rotate with its centre fixed.

{\bf 6. Conclusions}

In this paper, a time-dependent Ginzburg--Landau model for a
complex scalar field coupled to electromagnetism in two space 
dimensions has been considered. The Lagrangian
incorporates the standard potential energy of Ginzburg--Landau theory, and
there is a Schr\"odinger kinetic term
for the scalar field and a Chern--Simons term for the electromagnetic
field. The model is exactly Galilean invariant, so fields respond to a
transport current simply by a velocity boost parallel to the current.
The model has vortex solutions, and at the critical coupling $\lambda = 1$,
it has the $2n$-dimensional manifold $M^n$ of static $n$-vortex
solutions satisfying the Bogomolny equations. The parameters of these
solutions are the positions of the $n$ zeros of the scalar field, which can be
identified with the vortex positions. 

If $\lambda \ne 1$, and there are two or more vortices, then 
they generally move, but without dissipation.
For $\lambda$ close to $1$, the field dynamics describing vortex
motion can be approximated by considering solutions of the Bogomolny
equations with time-varying parameters. This leads to a reduced
Lagrangian for motion on $M^n$, with
a kinetic term linear in the vortex velocities, and a
potential term. Remarkably, the kinetic term depends only on local
data associated with each vortex. It defines a gauge potential
on $M^n$ which depends smoothly on the vortex 
positions, even as the vortices become coincident.

For two vortices, the gradient of the potential energy is along the
line joining them, but the motion is at right angles to this, and they
orbit each other at constant separation. It would be interesting to
study the motion of more than two vortices in this model. This 
will require more detailed
computations of the potential energy and the gauge potential
on $M^n$, or a numerical simulation of the field dynamics.

The quantisation of the vortex motion should also be considered.

\vskip 10pt
{\bf Acknowledgements}

I am grateful to several people for discussions about this work,
including N. Goldenfeld, D. Thouless, G. Volovik, A. Gill, and
J. Dziarmaga. I am especially indebted to I. Aitchison for his DAMTP seminar
on the time-dependent Ginzburg--Landau theory, and to 
G. Gibbons for discussions about Galilean invariance.

This work is partly supported by EPSRC grant GR/K50641, part of the
Applied Nonlinear Mathematics Programme.

\vfil\eject
{\bf References}

\item {1.} C.P. POOLE JR., H.A. FARACH AND R.J. CRESWICK,
``Superconductivity'', Academic Press, San Diego, 1995;

D.R. TILLEY AND J. TILLEY, ``Superfluidity and Superconductivity (3rd
ed.)'', Inst. of Phys. Publishing, Bristol, 1990.

\item {2.} A. ORAL, S.J. BENDING AND M. HENINI,
Appl. Phys. Lett. {\bf 69} (1996) 1324. 

\item {3.} A.A. ABRIKOSOV, Sov. Phys. JETP {\bf 5} (1957) 1174.

\item {4.} L. JACOBS AND C. REBBI, Phys. Rev. {\bf B19} (1979) 4486. 

\item {5.} H.B. NIELSEN AND P. OLESEN, Nucl. Phys. {\bf B61} (1973) 45. 

\item {6.} A. VILENKIN AND E.P.S. SHELLARD, ``Cosmic Strings and
other Topological Defects'', Cambridge U.P., Cambridge, 1994.

\item {7.} L.P. GOR'KOV AND G.M. \'ELIASHBERG, Sov. Phys. JETP {\bf 27}
(1968) 328;

S.J. CHAPMAN, S.D. HOWISON AND J.R. OCKENDON, SIAM Rev. {\bf 34}
(1992) 529.

\item {8.} S. DEMOULINI AND D. STUART, ``Gradient flow of the
superconducting Ginzburg--Landau functional on a plane'', to appear in
Commun. Anal. and Geom.;

D. STUART, Appl. Math. Lett. {\bf 9} (1996) 27.

\item {9.} P.G. DE GENNES, ``Superconductivity in Metals and Alloys'',
Benjamin, New York, 1966;

P. NOZI\`ERES AND W.F. VINEN, Philos. Mag. {\bf 14} (1966) 667.

\item {10.} I.J.R. AITCHISON, P. AO, D.L. THOULESS AND X.-M. ZHU,
Phys. Rev. {\bf B51} (1995) 6531;
 
M. STONE, Int. J. Mod. Phys. {\bf B9} (1995) 1359.

\item {11.} T.M. SAMOLS, Commun. Math. Phys. {\bf 145} (1992) 149.

\item {12.} N.S. MANTON, Phys. Lett. {\bf B110} (1982) 54.

\item {13.} P.A. SHAH, Nucl. Phys. {\bf B429} (1994) 259.

\item {14.} R. JACKIW AND S.-Y. PI, Phys.Rev. {\bf D42} (1990) 3500.

\item {15.} I.V. BARASHENKOV AND A.O. HARIN, Phys. Rev. Lett. {\bf 72}
(1994) 1575.

\item {16.} A.T. DORSEY, Phys Rev. {\bf B46} (1992) 8376.

\item {17.} M. LE BELLAC AND J.-M. L\'EVY-LEBLOND, Nuovo Cim. {\bf
14B} (1973) 217.

\item {18.} E.B. BOGOMOLNY, Sov. J. Nucl. Phys. {\bf 24} (1976) 449.

\item {19.} A. JAFFE AND C. TAUBES, ``Vortices and Monopoles'',
Birkh\"auser, Boston, 1980;

C. TAUBES, Commun. Math. Phys. {\bf 72} (1980) 277.

\item {20.} Q. LIU AND A. STERN, Phys. Rev. {\bf D52} (1995) 1300.

\item {21.} J. DZIARMAGA, Phys. Rev. {\bf B53} (1996) 8231.

\bye